# Chapter 29
# Beyond a pale blue dot : how to search for possible bio-signatures on earth-like planets

**Yasushi Suto**

Department of Physics and Research Center for the Early Universe

The University of Tokyo, Tokyo 113-0033, Japan

**Abstract** The Earth viewed from outside the Solar system would be identified merely like a pale blue dot, as coined by Carl Sagan. In order to detect possible signatures of the presence of life on a *second earth* among several terrestrial planets discovered in a habitable zone, one has to develop and establish a methodology to characterize the planet as something beyond a mere pale blue dot. We pay particular attention to the periodic change of the color of the *dot* according to the rotation of the planet. Because of the large-scale inhomogeneous distribution of the planetary surface, the reflected light of the *dot* comprises different color components corresponding to land, ocean, ice, and cloud that cover the surface of the planet. If we decompose the color of the dot into several principle components, in turn, one can identify the presence of the different surface components. Furthermore, the vegetation on the earth is known to share a remarkable reflection signature; the reflection becomes significantly enhanced at wavelengths longer than 760nm, which is known as a red-edge of the vegetation. If one can identify the corresponding color signature in a pale blue dot, it can be used as a unique probe of the presence of life. I will describe the feasibility of the methodology for future space missions, and consider the direction towards astrobiology from an astrophysicist's point of view.

**Key words**: bio-signatures, pale-blue-dot, red-edge, Copernican Principle

## 29. 1  Introduction

Discovery of an amazing number of exoplanetary systems since 1995 has completely changed our view of the world itself. In particular, we learned once again the universal validity of the Copernican Principle; we do not occupy any special place in the universe. Indeed, this is exactly the very important philosophical lesson that we have learned in the history of astronomy over and over again.

A straightforward corollary of the Copernican Principle is that our earth is simply just one of numerous planets in the universe that harbor the life. This will be easily expected from a very crude, order-of-magnitude argument shown below.

The mass of our Galaxy is approximately $10^{11} M_{sun}$, which implies that there are roughly $10^{11}$ stars (in the current argument, we neglect the dark matter contribution and the mass function of stars, and simply assume that the typical mass of stars is $1 M_{sun}$. This would change the result merely by a few orders of magnitude, and thus the final conclusion below is not affected at all!).

Current planet surveys have revealed that most stars host at least one planet, and that dozens out of several thousands of host stars, therefore roughly 0.1%, turn out to have more than one rocky planets located in a habitable zone (e.g., Kasting 1993, Kopparapu et al. 2013), i.e., the equilibrium temperature of the planet is between 0 and 100 degrees in Celsius. Thus, if H2O exists abundantly, it is expected to be liquid on the surface of the planet. The word "habitable" is quite misleading in a sense that the range of equilibrium temperature on the planet surface is supposedly neither a necessary nor sufficient condition for the existence of life. Furthermore, the existence of abundant water on those planets are not at all discovered observationally (yet). Nevertheless, it is a reasonable working hypothesis to proceed further here, and let me use "temperate" instead of "habitable" according to relatively recent literatures.

This implies that we would have $10^8$ temperate planets in our Galaxy. It should be emphasized that the value is estimated now from the observed facts. On the other hand, the relative fraction of planets, P, with a reasonable amount of water, and possibly with a reasonable amount of lands on the surface as well, is quite uncertain at this point, but will be estimated observationally in future by a remote sensing as described in section 29.3.

If a planet has a right amount of water for bearing life, what is the probability that the planet eventually develops life? This is intrinsically difficult and almost impossible question to answer scientifically. Therefore, we need to resort to the Copernican Principle. Our Solar system was born about 4.6 Gyr ago, and the first life on the earth is supposed to have emerged approximately 1Gyr later. Thus, the emergence of life itself may not be such a rare event as long as the relevant environment, which we do not yet understand exactly, is provided. A simple application of the Copernican Principle suggests that a fairly large fraction of temperate planets with oceans and lands will inevitably develop a certain type of life.

Plants on the earth went out of oceans and started to grow on lands about 0.5Gyr ago; from the view-point of remote sensing, this is the most important event in the evolution of life. It is not clear at all how long such planets continue to be detectable via remote-sensing, since we have no idea if any lifeform exhibiting a significant bio-signature survives possible drastic environment changes including astronomical impacts, geophysical activities, and human wars. Even if we assume pessimistically that our earth stops exhibiting detectable bio-signatures very soon, the fraction of detectable period of our earth via remote-sensing (of course, just in principle) over the life-time of the Sun would be 0.5Gyr/10Gyr=0.05.

Therefore, out of the $10^8$ temperate planets in our Galaxy, 5x$10^6$P would potentially exhibit some kinds of bio-signatures for remote-sensing. Again, on the basis of the Copernican Principle, P is unlikely to be sufficiently small, and we expect that $5 \times 10^6 P \gg 1$. This implies that we have to consider seriously how to search for possible bio-signatures on earth-like planets.

## 29.2 Lessons from previous pioneering attempts

Vest Melvin Slipher is a renowned astronomer, and is well known for his contribution to the discovery of redshifts of distant galaxies. Indeed, Edwin Hubble owed Slipher's measurements of galaxy redshifts in proposing his "famous" distance-redshift relation (Hubble 1929), which eventually led to the standard model of the expanding universe. It is now well recognized that Slipher's contribution to the discovery of the expansion of the universe has been significantly underestimated; I would definitely recommend Peacock (2013) for interested readers, which nicely describes numerous great achievements of Slipher.

Actually, his pioneering contribution to astrobiology seems to have been equally underestimated either. In his paper entitled "Observations of Mars in 1924 Made at Lowell Observatory II. Spectrum observations of Mars" (Slipher 1924), he attempted to test the existence of chlorophyll in the dark region on Mars. He clearly recognized the importance of the reflection spectrum feature of vegetation as a possible bio-signature on Mars. He noted that *"The reflection spectrum from vegetation is not at all definite visually as its most distinctive feature is its brilliancy in the deep red, beyond the sensitivity of the eye"*.

This characteristic feature, often referred to as the red-edge of vegetation, (sharp increase of reflection spectrum beyond around 0.75μm) is supposed to be very generic over most plants on Earth, and understood to be related to the efficiency of the photosynthesis.

He concluded his paper by stating that *"The Martian spectra of the dark regions so far do not give any certain evidence of the typical reflection spectrum of chlorophyl. The amount and types of vegetation required to make the effect noticeable is being investigated by suitable terrestrial exposures"*. I believe that this is quite amazing and pioneering work in the history of astrobiology.

Interestingly, there have been several observational claims of spectroscopic evidence for vegetation on Mars on the basis of a different absorption feature around 3.4μm. e.g., Sinton (1957) and also Briot, Schneider, and Arnold (2004) for a historical overview. Assuming that they are not reliable, Slipher's idea has been seriously considered and applied in the modern context for the first time by Sagan et al. (1993). They searched for bio-signatures on Earth at the first fly-by of Galileo space-craft on December 8, 1990, and successfully concluded that there is life on Earth!

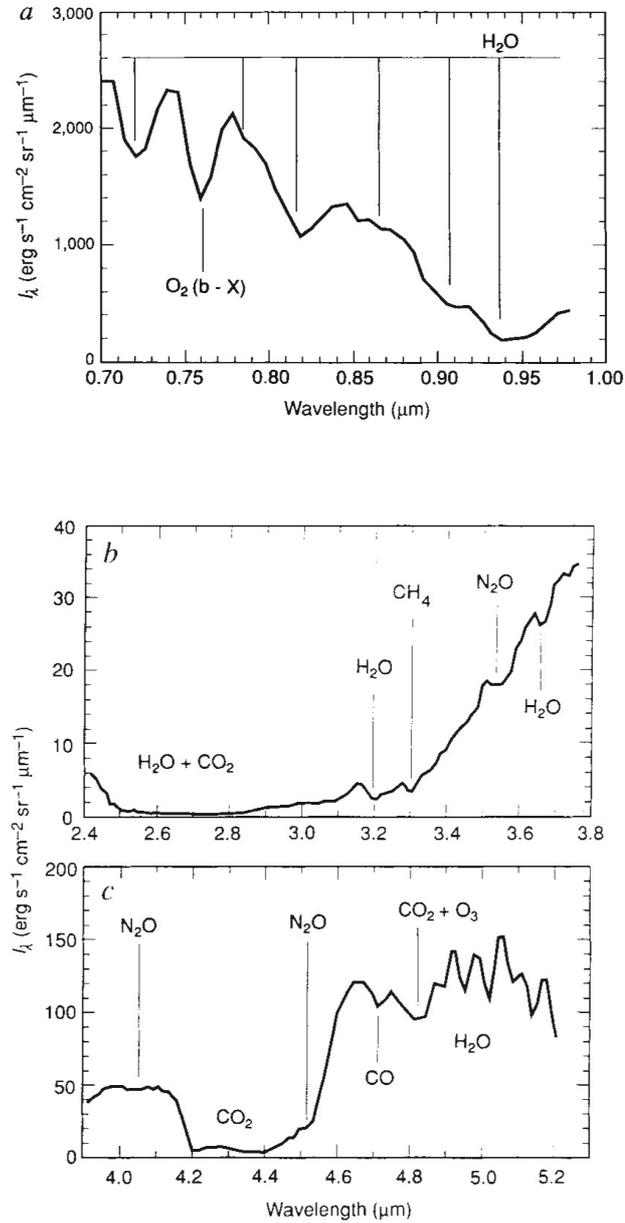

**Fig. 1** Spectra of Earth observed by Galileo spacecraft in December 1990 over a relatively cloud-free region of the Pacific Ocean. The molecules responsible for major absorption bands are indicated in each panel (reprinted from Figs.1a, 1b, and 1c of Sagan et al. 1993).

Bio-signatures that they detected from the remote-sensing of Earth include (i) abundant gaseous oxygen in visible and near-infra-red bands, (ii) atmospheric methane of the significantly larger abundance than expected from simple thermal equilibrium, and (iii) red-edge feature around 0.75μm.

Figure 1 plots the spectra of Earth over a relatively cloud-free region of the Pacific Ocean observed by Galileo (reprinted from Figure 1 in Sagan et al. 1993). Fig. 1a is the spectrum for $0.70 < \lambda[\mu m] < 1.0$, showing a strong absorption feature of the A band molecular oxygen at 0.76μm with several $H_2O$ absorptions as well. The column density of $O_2$ is estimated to be about 200 g cm$^2$ and such a large abundance is very unlikely to be produced and accumulated by any abiotic process like the UV photo-dissociation of water followed by the Jeans escape of hydrogen to space. Thus Sagan et al. (1993) concluded that *"Galileo's observations of $O_2$, thus at least raise our suspicions about the presence of life"*.

In the near-infrared bands, $2.4 < \lambda[\mu m] < 3.8$ (Fig.1b) and $3.9 < \lambda[\mu m] < 5.3$ (Fig.1b), a very strong $CO_2$ absorption band can be clearly identified around 4.3μm, as well as several $N_2O$ and $H_2O$ absorptions. Most notably, they identified the methane feature at 3.31μm, and found that the derived abundance is about 140 orders of magnitude higher than a simple expectation from thermal equilibrium. Since $CH_4$ is supposed to be oxidized quickly to $CO_2$ and $H_2O$, a continuous pumping source of $CH_4$ is required, which is most likely life. This is also the case for the high disequilibrium abundance of $N_2O$, which will be due to the presence of nitrogen-fixing bacteria and algae.

Such abundant atmospheric molecules can be used as important bio-signatures in classical astronomical observations, but still may not be directly related to the presence of life. Indeed the biological interpretation of the observed methane and other molecules may not be so robust; for instance, Epiope and Sherwood Lollar (2013) discussed the possible abiotic origin of methane in the planetary atmosphere. The detection of the red-edge feature, in turn, is challenging, but, if detected at all, would be interpreted as more straightforward evidence for the life dominating a fair fraction of the planet.

As shown in Figure 2 (reprinted from Figure 2c and Figure 3 in Sagan et al. 1993), Sagan et al. (1993) presented broad-band spectra of the three different regions on Earth, and argued that the unusually strong absorption features in the spectra of areas B and C are *"the signature of a light-harvesting pigment in a photosynthetic system"*, while that of area A is consistent with a variety of dark rock or mineral-soil surfaces.

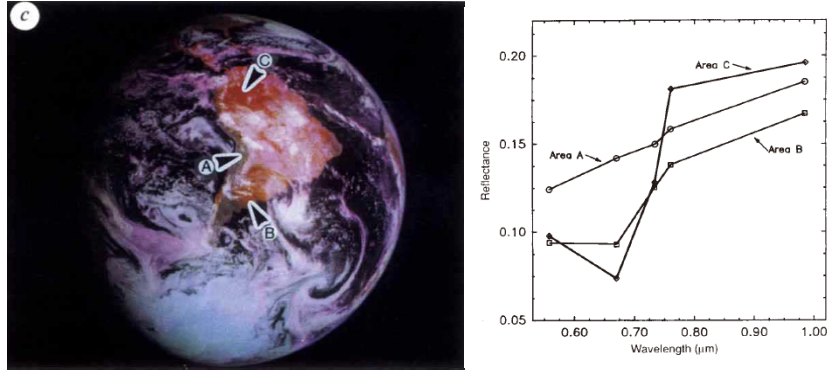

**Fig. 2** (Left) Composite color image of Earth observed by Galileo spacecraft in December 1990. (Right) Spectra corresponds to the three areas A, B, and C in the left panel (reprinted from Figs.2c and 3 of Sagan et al. 1993).

## 29. 3 Colors of a second earth

It is no doubt that Sagan et al. (1993) is the first serious observational attempt to present fundamental methodologies to search for life in planets from remote-sensing data. It is even more amazing to recognize that it was before the first discovery of an exoplanet around a sun-like star (Mayor and Queloz 1995). Unfortunately, however, their method is partially based on the spatially-resolved imaging observation of the surface of Earth, and thus is not directly applicable to exoplanets.

Ford, Seager, and Turner (2001) is the first to realize such a basic limitation. They computed diurnal photometric variability of Earth in different bands by averaging over the visible part illuminated by the Sun for a distant observer. In particular, they claimed that the photometric variability due to the red-edge feature may be marginally detectable for future space interferometer missions. The diurnal photometric variability is a more realistic approach with remote-sensing of earth-like exoplanets since it is based on the continuous monitoring of a change of colors of spatially "one dot". Indeed, this implies that Earth is not a mere *pale blue dot*, but a color-changing dot due to its spin-modulated surface landscape.

Inspired by this basic idea, we started a systematic study of the diurnal photometric variability of Earth (Fujii et al. 2011, 2012). We have not only computed the expected variability, but attempted to estimate the area fraction of different surface components including snow, land, ocean and vegetation by inverting the simulated light-curves in different photometric bands (see also Cowan et al. 2009, Majeau, Agol, and Cowan 2012, Fujii, Lusting-Yaeger, and Cowan 2017, and references therein).

To be more specific, we first created mock light curves for Earth *without clouds* in different photometric band, using empirical data from satellites. These light curves were attempted to be fit to an isotropic scattering model consisting of four surface types: ocean, soil, snow, and vegetation as shown in Figure 3. We considered a very idealized observational situation in which the light from the host star is completely blocked, and the photometric noise is due to the Poisson fluctuations in the observed photon counts from the planet alone.

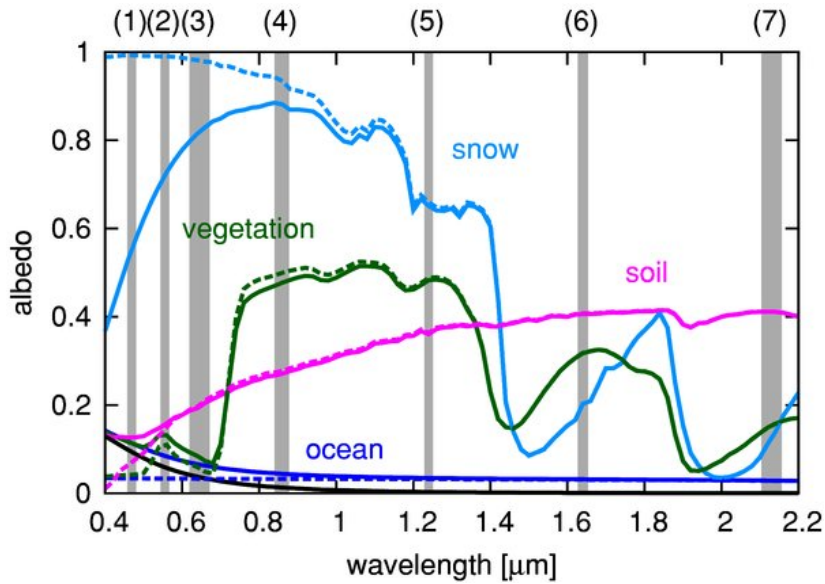

**Fig. 3** Wavelength-dependent effective albedos of ocean (blue), soil (magenta), vegetation (green), snow (cyan), and atmosphere with Rayleigh scattering alone (black). The solid lines show the effective albedo with Earth-like atmosphere, while the dashed lines show the effective albedo without an atmosphere. Shaded regions correspond to the MODIS (MODerate resolution Imaging Spectroradiometer) bands. The numbers at the top are the labels of the different photometric bands (reprinted from Figure 7 of Fujii et al. 2010).

Figure 4 presents an example of our decomposition of light-curves in terms of the four surface components using the photometric bands (1) to (5) indicated as gray bars in Figure 3. This simulated observation assumes an earth-twin from 10 pc away from us, and a dedicated space telescope of diameter 2m with an exposure time of 1 hour over 2-week continuous monitoring. In such an idealized situation where the light from the host star is completely blocked and the planetary surface is not covered by clouds, we are able to recover the correct the fractional areas of surface components fairly well. In particular, at a certain phase of Earth in which plants cover a large fraction of the planetary surface, we may be able to even detect the presence of vegetation via its distinct spectral feature of photosynthesis.

Admittedly our assumptions may not be so realistic. Especially, the significant cloud coverage would be inevitable for any temperate terrestrial planets with abundant liquid water.

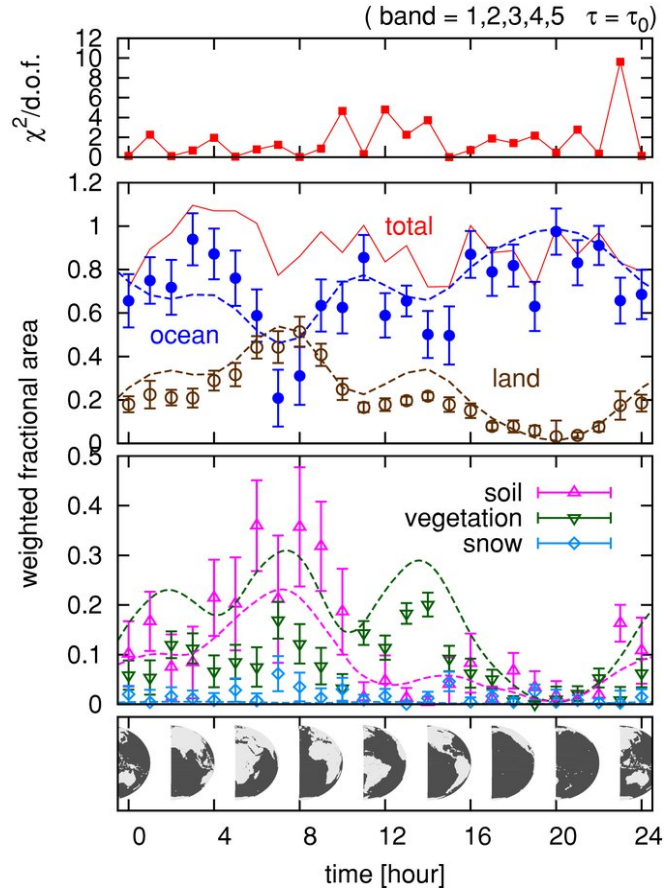

**Fig. 4** Reconstructed fractional areas for four surface types from the simulated light curves in five bands (bands 1 - 5). The top panel shows the value of reduced $\chi^2$ for each epoch. The upper middle panel displays the results of estimating weighted fractional areas of ocean (blue), land (=soil+vegetation+snow; brown), and the total of them (red). The lower middle panel displays those of soil (magenta), vegetation (green), and snow (cyan). The dashed lines in those two panels show the weighted fractional areas derived from the real classification dataset by the MODIS satellite. The quoted error bars indicate the variance of the best-fit values from 100 realizations. The bottom panel depicts the snapshots of the Earth at the corresponding epochs where the ocean is painted in gray and the land in white (reprinted from Figure 8 of Fujii et al. 2010).

Therefore, we considered the degree of degradation of our surface recovery method due to the cloud coverage by applying it to multi-band diurnal light curves of Earth from the EPOXI spacecraft. The detailed discussion can be found in Fujii et al. (2011), and here we simply show the resulting longitudinal map recovered from the diurnal variations in Figure 5. While the angular resolution is inevitably low, it is encouraging that some of the major geographical features of the Earth e.g. two oceans, the Sahara desert, and the two largest land masses can be approximately identified, even from the color-changing dot alone.

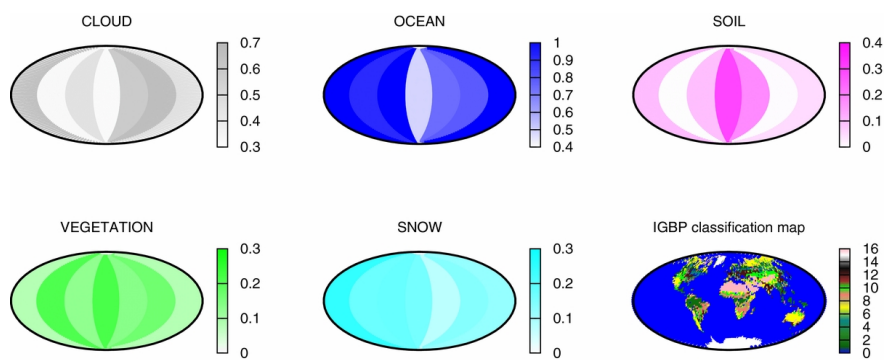

**Fig. 5** The 7-slice longitudinal distribution of each surface component recovered from the June light curves. The indices of the IGBP classification map (the bottom right panel: http://modis-atmos.gsfc.nasa.gov/ECOSYSTEM/index.html) are--- 0: ocean, 1: evergreen needleleaf forest, 2: evergreen broadleaf forest, 3: deciduous needleleaf forest, 4: deciduous broadleaf forest, 5: mixed forest, 6: closed shrubland, 7: open shrubland, 8: woody savannas, 9: savannas, 10: grasslands, 11: permanent wetlands, 12: croplands, 13: urban and built-up, 14: cropland/natual vegetation mosaic, 15: snow and ice, and 16: barren or sparsely vegetated (reprinted from Figure 16 of Fujii et al. 2011).

## 29. 4 Conclusion

Apparently, there remain many things to be improved in methodology both theoretically and observationally. Fujii et al. (2010, 2011) exploited the color modulation of the visible surface due to the planetary spin alone. If the planetary spin axis is oblique with respect to its orbital axis, as is the case with Earth, the resulting annual modulation may even allow the recovery of the two-dimensional surface map of the planet. This interesting possibility has been studied by Kawahara and Fujii (2010, 2011) and Fujii and Kawahara (2012), and they showed that it is possible to reproduce the 2D surface map of Earth, at least in an idealized situation. Kawahara (2016) also proposed novel methodology to determine the planetary obliquity from frequency modulations of photometric light-curves.

10Those preliminary feasibility studies have adopted simulated and/or observed datasets of Earth itself. Needless to say, it is by far the most important check to begin with, but it is also true that such studies never cover a possible range of realistic diversities of candidate planets. This is a fundamental and intrinsic limitation in considering bio-signatures, given the fact that we have no idea of life outside Earth. Nevertheless, it is such an important and fascinating topic, not only in astronomy and astrobiology, but even in all modern sciences. We need to continue, and definitely will be able to realize that *"We did not know anything"* (Asimov 1914).